\documentclass[epj]{svjour} 
\usepackage[mathscr]{eucal}
\usepackage{amsmath}
\usepackage{amssymb}
\usepackage{amsfonts}
\usepackage{ae}
\usepackage{epsfig}
\begin{document}

\def\BE{\begin{equation}}
\def\EE{\end{equation}}
\def\BY{\begin{eqnarray}}
\def\EY{\end{eqnarray}}

\def\L{\label}
\def\nn{\nonumber}
\def\({\left (}
\def\){\right)}
\def\[{\left [}
\def\]{\right]}
\def\o{\overline}
\def\BA{\begin{array}}
\def\EA{\end{array}}
\def\ds{\displaystyle}

\title{Quantum state of an injected TROPO above threshold : purity,
  Glauber function and photon number distribution}

\author{T. Golubeva\inst{1}, Yu. Golubev\inst{1}, C. Fabre\inst{2},
  N. Treps\inst{2}}
\institute{V.~A.~Fock Physics Institute,
  St.~Petersburg State University, 198504 Stary Petershof,
  St.~Petersburg, Russia \and Laboratoire Kastler Brossel, Université
  Pierre et Marie-Curie-Paris 6, ENS, CNRS ; 4 place Jussieu, 75005
  Paris, France}
\date{\today}

\abstract {In this paper we investigate several properties of the
full signal-idler-pump mode quantum state generated by a triply
resonant non-degenerate Optical Parametric Oscillator operating
above threshold, with an injected wave on the signal and idler
modes in order to lock the phase diffusion process. We determine
and discuss the spectral purity of this state, which turns out not
to be always equal to 1 even though the three interacting modes
have been taken into account at the quantum level. We have seen
that  the purity is essentially dependent on the weak
intensity of the injected light and on an asymmetry of the
synchronization. We then derive the expression of its total
three-mode Glauber P-function, and calculate the joint
signal-idler photon number probability distribution and
investigate their dependence on the injection.}


\authorrunning{T. Golubeva, Yu. Golubev, C. Fabre, N. Treps}
\titlerunning{Quantum state of an injected TROPO above threshold}

\maketitle

\section{Introduction}

Three-wave nonlinear interaction in a $\chi^{2}$ medium is one of the
main model systems of quantum optics. The problem is simplified when
one inserts an optical cavity around the non-linear medium, because
the resonances of the cavity permit to restrict the analysis to the
intracavity resonant modes only. The system is called in this case a
"TROPO" (Triply Resonant Optical Parametric Oscillator).  In addition,
in the bellow threshold, one has generally the possibility to consider
the pump as a coherent one with a fixed classical amplitude and thus
to restrict the analysis to a two-mode problem (idler and signal).  It
has been shown that such a system produces a squeezed vacuum state
below the oscillation threshold when the two quantum modes are
degenerate \cite{squeeze1}, and twin, quantum intensity correlated
beams above the threshold when they are
non-degenerate\cite{twin1}. Both predictions have been confirmed over
the last two decades by many experiments\cite{squeeze2,twin2}. It has
been also shown that this system has many other interesting quantum
properties, especially in the non-degenerate case: phase
anti-correlations\cite{squeeze3}, and EPR entanglement below and above
threshold\cite{EPR1}. This last feature has been recently confirmed
experimentally\cite{EPR1b,EPR2}.

In the above-threshold regime, where the three fields have intensities
which are of the same order of magnitude, the assumption that the
intracavity pump field is classical is not valid. One is faced with a
true three-quantum-mode problem. In such a system, not only a
conversion of the pump wave into the signal and idler waves, but a
mutual conversions of all three modes take place. It was theoretically
shown\cite{squeeze3} that under this conditions the pump wave turns
out to be non-classical too. It turns out to be significantly
squeezed in some parts of the parameter space, and this was confirmed
experimentally\cite{pompe2}. More recently it has been shown
that the pump beam is quantum-correlated with the sum of the signal
and idler fields, and that there is actually a strong three-partite
EPR entanglement between the three interacting
fields\cite{tripartite}. This shows that the TROPO, which has been one
of the most studied systems by quantum opticians for decades, can
still provide us with good surprises.

Moreover, in the recent years, the attention of the quantum optics
community has gradually shifted from the determination of squeezed
variances and quantum correlations of various observables of the
system to more subtle characterizations and manipulations of its
quantum state. There is for instance a strong development of
studies concerning states produced by conditional measurements
performed on continuous variable optical systems, such as photon
deletion techniques or selection of instantaneous values
of the photocurrent fluctuations. For such studies, one
needs to know the full quantum state of the system under
consideration, and not only the second order moment of the
observed quantities.

Whereas it is simple to define a quantum state of a confined
optical system, such as a cavity mode or a single light
pulse, it turns out to be very difficult to rigorously
define the quantum state of a c.w. optical beam propagating from
the generating optical device to the detectors, even though such a
concept is often used in the community from an intuitive point of
view. One difficulty that one encounters is that the state of the
system crucially depends on the exact measurement that one wants
to perform on it. For example, quantum properties are present in
the system when measured using long integration times, and the
system is perfectly classical for short integration times.

The purpose of this paper is to bring some insight to this
important issue by considering as an example the "toy-model" of
the light generated by a TROPO, and to present various ways of
characterizing its full quantum state. We will consider here only
the above threshold case, where all the generated fields have
significant mean values, so that the linearization method for
treating the fluctuations\cite{book,ProgressInOptics} is
undoubtedly valid except very close to the oscillation threshold.
As is well-known, as far as the signal-idler phase difference is
concerned, a phase diffusion phenomenon takes places, giving rise
to diverging phase difference fluctuations at very long
times\cite{Graham}. To eliminate this complication, we will assume
here that the TROPO is injected on the signal and/or idler modes,
which has the effect of synchronizing the two fields and locking
the phase diffusion effect. In this case, the validity of the
method is unquestionable, as well as the Gaussian character of all
the output fields.

As a starting point we will determine the full $6 \times 6$
covariance matrix for the spectral components of the quantum
fluctuations of the three output modes. This enables us to
calculate the spectral purity of the TROPO quantum state, and to
discuss in which respect the system can be described or not by a
three-field state vector. We then derive the expression of the
intracavity stationary Glauber function for the whole system. Its
interest lies in the fact that the quasi-probability function contains
in a condensed form all the quantum properties. The way it is
written, and the symmetries that it reveals, are a guide to
determine which are the combinations of the different modes which
have the best quantum properties. We use it to determine the
P-function and purity of the outfield state in a simple particular
case. We finally derive the joint signal-idler probability
distribution, a useful quantity to predict, using a quantum
state-reduction approach, the result of conditional measurements
performed on the system \cite{conditional1}.

The article is organized as follows. In section II, we give the
framework of the model that we use to describe the injected TROPO
above threshold. We then derive the expression of the Fourier
components of the quantum fluctuations of the three interacting
modes. In the next section, we determine and discuss the purity of
the system from the expression of the covariance matrix. We then
derive the Glauber stationary P function for the intracavity
fields. We use it to derive the expression of the Glauber
stationary P function for the output fields in the simple case
where the exposure time of the detection is small. In the last
section we derive and discuss the expression of the joint
signal-idler probability distribution as a function of the
detector exposure time. Details of the derivation in the
asymmetrical injection case are given in the appendix of the
paper.

\section{Physical model and general equations \L{II}}

\subsection{The master equation}

A non-degenerate optical parametric oscillator consists of a
non-linear $\chi^{(2)}$ medium inserted in a high-Q cavity. This
medium ensures the down-conversion process
$\omega_p\to\omega_i+\omega_s$ with exchange of a pump photon with
twin signal and idler photons. As is well known this interaction
can be described in the exact phase matching case by the effective
interaction Hamiltonian \cite{Graham,Gardiner}:
\BY
&&\hat V=i\hbar g\(\hat a_p\hat a_i^\dag \hat a_s^\dag-h.c.\)\\
\nonumber
&&\[\hat a_l,\hat a_l^\dag\]=1,\quad l=p,i,s \L{2.2}
\EY
In this approach, the three interacting intra-cavity modes are
quantized. The master equation in the interaction picture for
three-mode field density matrix $\hat\rho$ is :
\BY
&& \dot{\hat\rho}=-\frac{i}{\hbar}\[\hat V,\hat\rho\]-\hat
R\hat\rho+\sum\limits_{m=p,i,s}\hat D_m\hat\rho \L{2.}.
\EY
The operator $\hat R $ describes the damping of the intracavity
quantum oscillators and its action on the density matrix is
determined by the following equality in the case of exact
resonance between the pump, signal, idler fields and three cavity
modes:
\BY
&&\hat R\rho=\sum\limits_{m=p,i,s}\frac{\kappa_m}{2}\(\hat
a_m^\dag \hat a_m\hat \rho+\hat \rho \hat a_m^\dag \hat a_m -2
\hat a_m\rho \hat a_m^\dag\).\L{2.3}
\EY
where $\kappa_m$ is the energy damping rate of the $m$-mode. The
operators $\hat D_m$ ensure an excitation of each of the actual
modes by a quasi-classical or coherent state of amplitude
$\sqrt{N_m^{in}}$:
\BY
&&\hat D_m\;\hat\rho=\frac{\kappa_m}{2}\sqrt{N_m^{in}}\[ \hat
a_m^\dag-\hat a_m,\hat\rho\]
\EY
The external quasi-classical fields with a real amplitudes
$\sqrt{N_m^{in}}$ are in resonance with the m-waves.

The excitation of the idler and signal waves by the external
coherent waves is used to depress the phase diffusion in the
system. For this aim, it would be enough to put
$\sqrt{N_i^{in}}\equiv\sqrt{N^{in}}\neq0$ and $\sqrt{N_s^{in}}=0$.
However the asymmetry arising here  complicates both the
mathematical and physical situations. Much simpler
solutions take place as
$\sqrt{N_i^{in}}=\sqrt{N_s^{in}}\equiv\sqrt{N^{in}}$. In this
article we shall consider both models focusing in the body of the
article on the symmetrical case and giving all wished formulas for
the asymmetrical excitation in App.~\ref{A}.

We can derive from equation (\ref{2.}) the corresponding evolution
equation in the Glauber diagonal representation for the three
interacting modes. The Glauber function P is introduced by the
integral relation:
\BY
\nonumber \hat\rho(t)= \int\!\!\int\!\!\int
&&P(\alpha_p,\alpha_i,\alpha_s,t)
\;|\alpha_p,\alpha_i,\alpha_s\rangle\times\\
&&\langle\alpha_p,\alpha_i,\alpha_s|
d^2\alpha_p\;d^2\alpha_i\;d^2\alpha_s .\L{rho}
\EY
With symmetrical phase locking,
the master equation reads:
 \BY
&&\frac{\partial P(\alpha_p,\alpha_i,\alpha_s,t)}{\partial
t}=\sum\limits_{m=i,s}\frac{\kappa_m}{2}\frac{\partial
}{\partial\alpha_m}(\alpha_m-\sqrt{N^{in}})P+\nn\\
&&+\frac{\kappa_p}{2}\frac{\partial
}{\partial\alpha_p}(\alpha_p-\sqrt{N_p^{in}})
P+g(\alpha_i\alpha_s\frac{\partial
P}{\partial\alpha_p}-\alpha_p\alpha_s^\ast\frac{\partial
P}{\partial\alpha_i}-\nn\\
&&-\alpha_p\alpha_i^\ast\frac{\partial
P}{\partial\alpha_s})+g\alpha_p\frac{\partial^2
P}{\partial\alpha_i\partial\alpha_s}+c.c.\L{8.}
\EY
The Glauber representation is often un-practical, as, when
non-classical effects are present, the Glauber functions can be
expressed only in terms of distributions, difficult to handle
mathematically. Sudarshan has suggested some way for writing these
distributions \cite{Sudarshan,Glauber} in the form of the momentum
series, however the master equation remains often impossible to
express simply. For instance, for the simplest model of the
sub-Poissonian laser \cite{Golubev}, the master equation contains
derivatives of all orders with respect to complex amplitudes,
which means that all momenta are connected with each other in one
system of equations of infinite order. In our problem, the
situation happens to be very favorable as the master equation has
a simple and well defined expression.

\subsection{Classical equations for the OPO operation}
As is well-known, to get the semi-classical evolution equations for
the mean amplitudes, we can use Eq (\ref{8.}), neglecting the second
derivatives with respect to the complex amplitude. Then one can get:
\BY
&&\dot\alpha_p=-\frac{\kappa_p}{2}\;(\alpha_p-\sqrt{N_p^{in}})-g\alpha_i\alpha_s\\
&&\dot\alpha_i=-\frac{\kappa_i}{2}\;(\alpha_i-\sqrt{N^{in}})+g\alpha_p\alpha_s^\ast\\
&&\dot\alpha_s=-\frac{\kappa_s}{2}\;(\alpha_s-\sqrt{N^{in}})+g\alpha_p\alpha_i^\ast
\EY
In the following we will only consider the case of equal losses
for the signal and idler cavity losses:
\BY
&&\kappa_i=\kappa_s\equiv\kappa.
\EY
Then it is not difficult to obtain the stationary solutions, which
can be written in the form:
 \BY
&&\alpha_p=\sqrt{N_p},\qquad \alpha_i=\alpha_s=\sqrt{N},
\EY
where for the values  $N$ and $N_p$ the following equalities take
place:
\BY
&& g\sqrt{N_p}=\frac{\kappa}{2}(1-\mu),\qquad
\frac{gN}{\sqrt{N_p}}=\frac{\kappa_p}{2}(\mu_p-1),\nn\\
&& \kappa(1-\mu)N=\kappa_p(\mu_p-1)N_p.
\EY
These equations depend on two dimensionless parameters $\mu_{p}$ and
$\mu$. $\mu_{p}$, the pump parameter, is defined as
\BY
&&\mu_{p}=\sqrt{\frac{N_p^{in}}{N_{th}}}
\EY
where $N_{th}=\kappa^2(1-\mu)/(4g^2)$. $\mu_{p}>1$ corresponds to
the above threshold regime. $\mu$, the injection parameter,
represents which fraction of the total signal and/or idler fields
is injected :
\BY
&& \mu=\sqrt{\frac{N^{in}}{N}}.
\EY
We will restrict our analysis to $\mu\ll 1$. There are two reasons
for this.  First, we are going to apply the limit of small photon
number fluctuation that is impossible in the bistability area. Second,
for a strong injected field, the Poissonian statistics of the
external field would be imposed on the intracavity mode and its
quantum properties would be destroyed.

\subsection{Limit of small amplitude and phase fluctuations}
We make now two important assumptions. First of all, because we
consider the above threshold situation inside a high-Q cavity, the
limit of the small photon number fluctuations can be used in our
treatment. If we present the complex amplitudes via amplitudes and
phases
\BY
&&\alpha_m=\sqrt{u_m}\;e^{\displaystyle i\varphi_m},\qquad
m=p,i,s,
\EY
then we can require
 \BY
&&u_m=N_m+\varepsilon_m,\qquad \varepsilon_m\ll N_m.
\EY
Secondly, because of injection, one can find well defined values
for the steady state phases, namely $\varphi_{m}=0$. We will
assume that in the system the phase fluctuations around these
steady state values are small :
 \BY
&&\varphi_m\ll1.
\EY
Taking into account this limit of small amplitude and phase
fluctuations, one finds that it is possible to factorize the
Glauber distribution in the form:
\BY
P(\alpha_p,\alpha_i,\alpha_s,t)&&=\\ \nn
&& =P(\varepsilon_p,\varepsilon_+,t)
P(\varepsilon_-,t)P(\varphi_p,\varphi_+,t) P(\varphi_-,t)
\EY
with decoupled equations for the different factors:
\BY
\frac{\partial P(\varepsilon_p,\varepsilon_+,t)}{\partial
t}&&=\left(\frac{1}{2}\frac{\partial}{\partial\varepsilon_p}(\kappa_p\varepsilon_p
+\kappa(1-\mu)\varepsilon_+)
+\right.\nn\\
&&+\frac{\partial}{\partial\varepsilon_+}\(\frac{\mu\kappa}{2}\varepsilon_+
-\kappa_p(\mu_p-1)\varepsilon_p\)+\nn\\
&&\left.+\kappa
N(1-\mu)\frac{\partial^2}{\partial\varepsilon_+^2}\right)P(\varepsilon_p,\varepsilon_+,t),\L{23.}\\
\frac{\partial P(\varepsilon_-,t)}{\partial
t}&&=\left(\kappa(1-\mu/2)\frac{\partial}{\partial\varepsilon_-}\varepsilon_--\right.\nn\\
&&\left. -\kappa N(1-\mu)
\frac{\partial^2}{\partial\varepsilon_-^2}\right)P(\varepsilon_-,t),\\
\frac{\partial P(\varphi_p,\varphi_+,t)}{\partial
t}&&=\left(\frac{\kappa_p}{2}\frac{\partial}{\partial\varphi_p}
(\varphi_p+(\mu_p-1)\varphi_+)+\right.\nn\\
&&+\frac{\partial}{\partial\varphi_+}
(\kappa(1-\mu/2)\varphi_+-\kappa(1-\mu)\varphi_p)-\nn\\
&&\left.-\frac{\kappa}{4N}(1-\mu)\;
\frac{\partial^2}{\partial\varphi_+^2}\right)P_(\varphi_p,\varphi_+,t),\L{}
\EY
\BY
&&\frac{\partial P(\varphi_-,t)}{\partial
t}=\(\frac{\mu\kappa}{2}\frac{\partial}{\partial\varphi_-}
\varphi_-+
\frac{\kappa}{4N}(1-\mu)\;\frac{\partial^2}{\partial\varphi_-^2}\)P(\varphi_-,t).\nn\\
&&\L{26.}
\EY
Here
\BY
\varepsilon_{\pm}=\varepsilon_i \pm \varepsilon_s,\quad
\varphi_{\pm}=\varphi_i \pm \varphi_s
\EY
One sees that the photon and phase fluctuations turn out to be
statistically independent at exact triple resonance. Furthermore,
the last equation shows that the injected fields lead to the
locking of the differential phase and to the suppression of the
phase diffusion phenomenon present in the degenerate OPO above
threshold. However the synchronizing fields influence the noise
properties of the system. It is important to understand whether
the phase locking and the presence of significant noise reduction
and quantum correlation are compatible with each other or not.
This is what we will see in the following.

\subsection{Intracavity spectral densities}
In order to analyze the time dependent correlation function, let
us derive the Langevin equations  for the three interacting fields
(\ref{23.})-(\ref{26.}). They are easily written according to
well-known rules and have the following form:
\BY
&&\dot\varepsilon_p=-\kappa_p/2\;\varepsilon_p-\kappa/2(1-\mu)\;\varepsilon_+,\\
&&\dot\varepsilon_+=-\kappa\mu/2\;\varepsilon_++\kappa_p(\mu_p-1)\;\varepsilon_p+f_+(t),\\
&&\dot\varepsilon_-=-\kappa(1-\mu/2)\;\varepsilon_-+f_-(t),\\
&&\dot\varphi_p=-\kappa_p/2\;\varphi_p-\kappa_p/2(\mu_p-1)\;\varphi_+,\\
&&\dot\varphi_+=-\kappa(1-\mu/2)\;\varphi_++\kappa(1-\mu)\;\varphi_p+g_+(t),\\
&&\dot\varphi_-=-\kappa\mu/2\;\varphi_-+g_-(t),
\EY
where the stochastic sources are determined by the pair
correlation functions
\BY
&&\langle f_+(t)f_+(t^\prime)\rangle=2\kappa N(1-\mu)\;\delta(t-t^\prime),\\
&& \langle f_-(t)f_-(t^\prime)\rangle=-2\kappa N(1-\mu)
\;\delta(t-t^\prime),\\
&&\langle g_+(t)g_+(t^\prime)\rangle=-\kappa(1-\mu) /(2N)\;\delta(t-t^\prime),\\
&& \langle g_-(t)g_-(t^\prime)\rangle=\kappa(1-\mu)/(2
N)\;\delta(t-t^\prime).
\EY
The best way to solve these equations is to rewrite them in the
Fourier domain and solve the simple algebraic system of equations
for the spectral components. One obtains
\BY
&&(\varepsilon_{+}^2)_\omega=2N\times\\
&&\times\frac{\kappa(\kappa_p^2+4\omega^2)(1-\mu)}
{\[2\omega^2-\kappa_p\kappa[\mu/2+(\mu_p-1)(1-\mu)]\]^2+\omega^2(\kappa_p+\kappa\mu)^2},\L{41.}\nn\\
&&(\varepsilon_{p}^2)_\omega=2N_p\times\\
&&\times\frac{\kappa^2\kappa_p(1-\mu)^2(\mu_p-1)}
{\[2\omega^2-\kappa_p\kappa[\mu/2+(\mu_p-1)(1-\mu)]\]^2+\omega^2(\kappa_p+\kappa\mu)^2},\nn\\
&&(\varepsilon_{p}\varepsilon_{+})_\omega=-2N\times\\
&&\times\frac{\kappa^2\kappa_p(1-\mu)}
{\[2\omega^2-\kappa_p\kappa[\mu/2+(\mu_p-1)(1-\mu)]\]^2+\omega^2(\kappa_p+\kappa\mu)^2},\nn\\
&&(\varepsilon_{-}^2)_\omega=-2N\;\frac{\kappa(1-\mu)}{\kappa^2(1-\mu/2)^2+\omega^2
},\L{44.}
\EY
and
\BY
&&(\varphi_{p}^2)_\omega=-\frac{1}{2N_p}\times\\
&&\times\frac{\kappa^2\kappa_p(\mu_p-1)(1-\mu)^2}
{\[2\omega^2-\kappa\kappa_p[\mu/2+\mu_p(1-\mu)]\]^2+
\omega^2[\kappa_p+2\kappa(1-\mu/2)]^2},\nn\\
&&(\varphi_{+}^2)_\omega=-\frac{1}{2N}\times\\
&&\times\frac{\kappa(\kappa_p^2+4\omega^2)(1-\mu)}
{\[2\omega^2-\kappa\kappa_p[\mu/2+\mu_p(1-\mu)]\]^2+
\omega^2[\kappa_p+2\kappa(1-\mu/2)]^2},\nn\\
&&(\varphi_{p}\varphi_{+})_\omega=\frac{1}{2N}\times\\
&&\times \frac{\kappa\kappa_p^2(\mu_p-1)(1-\mu)}
{\[2\omega^2-\kappa\kappa_p[\mu/2+\mu_p(1-\mu)]\]^2+
\omega^2[\kappa_p+2\kappa(1-\mu/2)]^2},\nn\\
&&(\varphi_{-}^2)_\omega=\frac{1}{2
N}\;\frac{\kappa(1-\mu)}{(\kappa\mu/2)^2+\omega^2 }.\L{48}
\EY
In these expressions, the spectral density
$(\varepsilon_{m}^2)_\omega$ and $(\varphi_{m}^2)_\omega$ are
defined as a factor in front of the delta-functions in the
correlation functions
\BY
&&\langle\varepsilon_{m}(\omega)\;\varepsilon_{m}(\omega^\prime)\rangle
=(\varepsilon_{m}^2)_\omega\;\delta(\omega+\omega^\prime),\qquad
m=p,\pm,\nn\\
&&\langle\varphi_{m}(\omega)\;\varphi_{m}(\omega^\prime)\rangle
=(\varphi_{m}^2)_\omega\;\delta(\omega+\omega^\prime),
\EY
where
\BY
&&\varepsilon_{m}(\omega)=\frac{1}{\sqrt{2\pi}}\int\varepsilon_{m}(t)\;e^{\displaystyle-i\omega
t}dt,\nn\\
&&\varepsilon_{m}(t)=\frac{1}{\sqrt{2\pi}}\int\varepsilon_{m}(\omega)\;e^{\displaystyle
i\omega t}d\omega.
\EY
The mutual correlations $(\varepsilon_{p}\varepsilon_{+})_\omega$
and $(\varphi_{p}\varphi_{+})_\omega$ are defined exactly in the
same way:
\BY
&&\langle\varepsilon_{+}(\omega)\;\varepsilon_{p}(\omega^\prime)\rangle
=(\varepsilon_{+}\varepsilon_{p})_\omega\;\delta(\omega+\omega^\prime),
\nn\\
&&\langle\varphi_{+}(\omega)\;\varphi_{p}(\omega^\prime)\rangle
=(\varphi_{+}\varphi_p)_\omega\;\delta(\omega+\omega^\prime).
\EY

\section{Spectral purity of the OPO quantum state \L{III}   }
\subsection{Output field variances and correlations}
In the previous section, we have considered the intracavity
spectral densities. Now we want to determine the corresponding
quantities for the output beams. We consider the optimum case
where only one mirror of the cavity is not perfectly reflecting
and transmits the light onto the detectors. Inside the cavity the
normalized amplitude was defined by the photon operators $\hat
a_m(t)\;(m=p,i,s)$ that obey the commutation relations
\BY
&&\[\hat a_m(t),\hat a_m^\dag(t)\]=\delta_{mn}.
\EY
Let us call $\hat A_m(t)$ the corresponding operator for the
output beams. They obey the commutation relations:
\BY
&&\[\hat A_m(t),\hat
A_n(t^\prime)^\dag\]=\delta_{mn}\delta(t-t^\prime),\nn\\
&&
\[\hat A_m(t),\hat A_n(t^\prime)\]=0.\L{k}
\EY
and are related to the intracavity ones by the input-output
relations on the coupling mirror:
\BY
&&\hat A_m(t)=\sqrt\kappa_m\;\hat a_m(t)-\(C_m+\hat
A_{m,vac}(t)\),\nn\\
&& m=p,i,s.\L{57}
\EY
We have taken into account the fact that the reflection
coefficient of the output mirror is about one. $C_p$ is the
complex normalized amplitude of the pump wave in resonance with
the p-mode. This amplitude is related to the quantity $N_p^{in}$
introduced earlier by relation $C_p=\sqrt{\kappa_pN_p^{in}}/2$.
The amplitudes $C_s$ and $C_i$ are the classical injected fields
in resonance with the idler and signal modes:
$C_i=C_s=\sqrt{\kappa N^{in}}/2$.

$\hat A_{m,vac}(t)$ are the input vacuum fluctuations, with
commutation relations
\BY
&&\[\hat A_{m,vac}(t),\hat
A_{n,vac}(t^\prime)^\dag\]=\delta_{mn}\delta(t-t^\prime),\nn\\
&&
\[\hat A_{m,vac}(t),\hat A_{n,vac}(t^\prime)\]=0.
\EY
We can rewrite (\ref{57}) for the fluctuations $\delta\hat
A_m=\hat A_m-\langle\hat A_m\rangle$ and $\delta\hat a_m=\hat
a_m-\langle\hat a_m\rangle$,
 in the Fourier domain:
\BY
&&\delta\hat A_m(\omega)=\sqrt\kappa_m\;\delta\hat
a_m(\omega)-\hat A_{m,vac}(\omega).\L{48.}
\EY
Let us divide the frequency scale into small equal intervals of
size $\Delta$. Then for each discrete frequency $\omega_l$ one can
write:
\BE
\delta\hat A_{m}^l=\frac{1}{\sqrt{\Delta}}
\int\limits_{\omega_l-\Delta/2}^{\omega_l+\Delta/2}\delta\hat
A_m(\omega)\;d\omega.
\EE
It is easy to check that the algebra of these operators is
determined by the commutation relations:
\BE
\[\delta\hat A_{m}^l,(\hat A_{m}^k)^\dag\]=\delta_{lk}.
\EE
Thus the operators $\delta\hat A_{m}^l,\;(\hat A_{m}^l)^\dag$ can
be thought of as being the annihilation and creation operators for
the photons in the frequency band around $\omega_l$. Then for each
mode (frequency) one can introduce the corresponding quadrature
components via the standard relations
\BY
&&\delta\hat X_{m}^l=\frac{1}{2}\((\delta\hat
A_{m}^l)^\dag+\delta\hat A_{m}^l\),\nn\\
&&\delta\hat Y_{m}^l=\frac{i}{2}\((\delta\hat
A_{m}^l)^\dag-\delta\hat A_{m}^l\).
\EY
Taking into account (\ref{48.}) the variances of the output
quadratures can then be expressed in terms of the intracavity
ones:
\BY
&&\langle\delta\hat
X^2_{m}\rangle_l=\frac{1}{4}+\frac{\kappa_m}{4N_m}\;\frac{1}{\Delta}
\int\limits_{\omega_l-\Delta/2}^{\omega_l+\Delta/2}(\varepsilon_m^2)_\omega\;d\omega,
\nn\\
&&\langle\delta\hat
Y^2_{m}\rangle_l=\frac{1}{4}+\kappa_mN_m\;\frac{1}{\Delta}
\int\limits_{\omega_l-\Delta/2}^{\omega_l+\Delta/2}(
\varphi_m^2)_\omega\;d\omega.
\EY
Going to the continuous frequencies by means of $\Delta\to0$,
one can obtain the wished expression giving the observed variances
and correlation in function of the intracavity ones:
 \BY
&&4(\delta\hat
X_m^2)_{\omega}=1+\kappa_m/N_m\;(\varepsilon_m^2)_{\omega},\nn\\
&&4(\delta\hat
Y_m^2)_{\omega}=1+4\kappa_mN_m\;(\varphi_m^2)_{\omega}.\L{54}
\EY
\BY
&&4(\delta\hat X_m\delta\hat
X_n)_{\omega}=\sqrt{\kappa_m/N_m}\sqrt{\kappa_n/N_n}\;(\varepsilon_m\varepsilon_n)_{\omega},\nn\\
&&4(\delta\hat Y_m\delta\hat
Y_n)_{\omega}=4\sqrt{\kappa_mN_m}\;\sqrt{\kappa_nN_n}\;(\varphi_m\varphi_n)_{\omega},\nn\\
&& m,n=p,i,s.\L{65}
\EY

\subsection{Spectral purity}

Let us now introduce the three-mode spectral covariance matrix of
the OPO. In the present
case of exact cavity resonance for the three modes, the amplitude
and phase fluctuations are independent and this matrix turns out
to be quasi-diagonal:
\BY
{\cal M}_{\omega}=\left(%
\begin{array}{cc}
  {\cal M}_{\epsilon} & 0  \\
  0 & {\cal M}_{\phi}  \\
  \end{array}
\right),
\EY
where $||{\cal M}_{\epsilon}||$ are amplitude  and $||{\cal
M}_{\phi}||$ are phase $3\times3$ matrices given by:
\BY
{\cal M}_{\epsilon}=\left(%
\begin{array}{ccc}
 4(\delta X_p^2)_\omega& 4(\delta X_p\delta X_i)_\omega &4(\delta X_p\delta X_s)_\omega \\
 4(\delta X_i\delta X_p)_\omega &4(\delta X_i^2)_\omega &4(\delta X_i\delta X_s)_\omega  \\
 4(\delta X_s\delta X_p)_\omega &4(\delta X_s\delta X_i)_\omega &4(\delta X_s^2)_\omega  \\
 \end{array}
\right),
\EY
and analogically $||{\cal M}_{\phi}||$ with help of the
variances\\
 $4(\delta Y_m\delta Y_n)_\omega$.
 The spectral variances are determined as a
factor in front of the delta-function in the correlation relation:
\BY
&&\langle\delta X_m(\omega)\delta
X_n(\omega^\prime)\rangle=(\delta X_m\delta
X_n)_\omega\;\delta(\omega+\omega^\prime),\nn\\
&&\langle\delta Y_m(\omega)\delta
Y_n(\omega^\prime)\rangle=(\delta Y_m\delta
Y_n)_\omega\;\delta(\omega+\omega^\prime)
\EY
A widely used way of describing the properties of a quantum system
at a given noise frequency $\omega$ is to assign to it a {\it
quantum state}, that we will call  "single noise frequency state",
described by the density matrix $\hat\rho_{\omega}$, that can be
experimentally characterized for example by quantum tomography.
Let us introduce the {\it spectral purity}, given by

\BY
&&{\Pi}_{\omega}=Tr\hat\rho_{\omega}^2.
\EY
This state is a pure state when
${\Pi}_{\omega}=1$, or a statistical mixture when ${\Pi}_{\omega}<1$. In the present case of small fluctuations and
input Gaussian states, all quantum fluctuations have Gaussian
statistics. Then the purity is equal to
\BY
&&{\Pi}_{\omega}=\frac{1}{\sqrt{\det {\cal M}_{\omega}}}.
\EY

\subsection{Spectral purity for a symmetrical injection}

In this simple case, we will use the basis of sum and difference
modes $m,n=p,\pm$ to calculate the determinant and therefore the
purity. The determinant of the covariance matrix can then be
written as the product:

\BY
&&\det{\cal M}_\omega={\cal N }_-{\cal D }_{X}{\cal D }_{Y},
\EY
where
\BY
&&{\cal N }_-=2\langle\delta\hat
X^2_{-}\rangle_{\omega}2\langle\delta\hat
Y^2_{-}\rangle_{\omega},\nn\\
&&{\cal D }_{X}=2\langle\delta\hat
X^2_{+}\rangle_{\omega}4\langle\delta\hat
X^2_{p}\rangle_{\omega}-4\langle\delta\hat X_{+}\delta\hat
X_{p}\rangle_{\omega}^2,\nn\\
&&{\cal D }_{Y}=2\langle\delta\hat
Y^2_{+}\rangle_{\omega}4\langle\delta\hat
Y^2_{p}\rangle_{\omega}-4\langle\delta\hat Y_{+}\delta\hat
Y_{p}\rangle_{\omega}^2,
\EY
One can see that ${\cal N }_-$ is related to the uncertainty
relation for the differential variances. ${\cal D }_{X,Y}$ is the
determinant of the $2\times2$ covariance matrix for the correlated
sum and pump amplitude and phase variances.

In this subsection we are going to consider, as an example and for the
simplicity of the calculation, the case $\kappa
(\mu_p-1),\kappa\ll\kappa_p$. According to the previous subsection the
differential variances are derived in the form
\BY
2\langle\delta\hat
X^2_{-}\rangle_{\omega}&&=1+\frac{\kappa}{2N}(\varepsilon^2_-)_\omega=\nn\\
&&= 1-\frac{\kappa^2(1-\mu)}{\kappa^2(1-\mu/2)^2+\omega^2 },\L{68}
\EY
\BY
&&2\langle\delta\hat Y^2_{-}\rangle_{\omega}=1+2\kappa
N(\varphi^2_-)_\omega=1+
\frac{\kappa^2(1-\mu)}{(\kappa\mu/2)^2+\omega^2 }.\L{69}
\EY
From these expressions, one retrieves first a striking but
well-known result, characteristic of the non-degenerate OPO above
threshold: ${\cal N }_-=1$ for any values of the parameters such
as the frequency, $\mu_p$, and $\mu$.

In order to get the other determinants ${\cal D }_{X,Y}$ we need
to derive the corresponding variances in the explicit form:
\BY
2\langle\delta\hat
X^2_{+}\rangle_{\omega}&&=1+\frac{\kappa}{2N}(\varepsilon^2_+)_\omega=\nn\\
&&=
1+\frac{\kappa^2(1-\mu)}{\kappa^2[\mu/2+(1-\mu)(\mu_p-1)]^2+\omega^2
},\L{70}\\
2\langle\delta\hat Y^2_{+}\rangle_{\omega}&&=1+2\kappa
N(\varphi^2_+)_\omega=\nn\\
&&=1-
\frac{\kappa^2(1-\mu)}{\kappa^2[\mu/2+(1-\mu)\mu_p]^2+\omega^2
},\L{71}\\
4\langle\delta\hat
X^2_{p}\rangle_{\omega}&&=1+\frac{\kappa_p}{N_p}(\varepsilon^2_p)_\omega=\nn\\
&&=
1+\frac{2\kappa^2(\mu_p-1)(1-\mu)^2}{\kappa^2[\mu/2+(1-\mu)(\mu_p-1)]^2+\omega^2
},\L{72}\\
4\langle\delta\hat Y^2_{p}\rangle_{\omega}&&=1+4\kappa_p
N_p(\varphi^2_p)_\omega=\nn\\
&&=1-
\frac{2\kappa^2(1-\mu)^2(\mu_p-1)}{\kappa^2[\mu/2+(1-\mu)\mu_p]^2+\omega^2
},\L{73}
\EY
\BY
2\langle\delta\hat X_{p}\delta\hat
X_{+}\rangle_{\omega}=&&\sqrt{\frac{\kappa_p}{2N_p}}\sqrt{\frac{\kappa}{2N}}\;
(\varepsilon_p\varepsilon_+)_\omega=\nn\\
=&&-\frac{\kappa^2\sqrt{(1-\mu)(\mu_p-1)}}
{\kappa^2[\mu/2+(\mu_p-1)(1-\mu)]^2+\omega^2}, \L{74}\\
2\langle\delta\hat Y_{p}\delta\hat
Y_{+}\rangle_{\omega}=&&\sqrt{2\kappa_pN_p}\sqrt{2\kappa
N}\;(\varphi_p\varphi_+)_\omega=\nn\\
=&&\frac{\kappa^2\sqrt{(\mu_p-1)(1-\mu)^3}}
{\kappa^2[\mu/2+\mu_p(1-\mu)]^2+ \omega^2}.\L{75}
\EY
Let us remark that, in contrast to ${\cal N }_-$, the quantities
\BY
&&{\cal N }_+=2\langle\delta\hat
X^2_{+}\rangle_{\omega}\;2\langle\delta\hat
Y^2_{+}\rangle_{\omega},\nn\\
&&{\cal N }_p=4\langle\delta\hat
X^2_{p}\rangle_{\omega}\;4\langle\delta\hat
Y^2_{p}\rangle_{\omega}
\EY
are non minimum, especially near zero frequencies when the pump is
not strongly above threshold. For example, when
$\mu_p-1\gg\mu $ and $\omega=0$
\BY
&&{\cal N }_+={\cal N
}_p=\frac{(\mu_p-1)^2+1}{\mu_p^2}\;\frac{\mu_p+1}{\mu_p-1}
\EY
${\cal N }_+$ and ${\cal N }_p$ take the minimum value compatible
with the Heisenberg inequality when the OPO is strongly above
threshold, but take very large values close to threshold.

Figure 1a gives the spectral purity of the three-mode OPO for
different values of the injection parameter ($\mu=0,1$) and of the
pump parameter ($\mu_p =4,2$ and 1,1). One first
observes that the spectral purity is equal to one outside the cavity
bandwidth in all configurations, and also well above threshold at
all frequencies. In contrast, the spectral purity at zero
frequency turns out to be less than one near threshold ($\mu_p
=1,1$).

We are therefore led to the conclusion that, {\it close to
threshold, the three-mode "single noise frequency state"
describing the OPO is mixed at low noise frequencies}. Considering
that we have taken into account in a quantum way all the
interacting modes, and that all the intracavity modes are
transmitted onto the detectors without any losses, so that the
total input-output evolution matrix is unitary, this result is
somewhat unexpected. But we must recall that we discuss here the
purity not of the complete system, but the purity of some of its
spectral components. Obviously, the intracavity parametric
interaction does not change the purity of the system as a whole.
However it leads to a redistribution of fluctuations and
correlations between the spectral components of the pump, signal
and idler modes, and as a result the purity of each spectral
component does not generally survive. Such a redistribution is
maximum when the coupling is maximum, i.e. at zero noise
frequencies, and vanishes outside the cavity bandwidth. Figure 1b
shows the spectral purity for larger injection. One can see that in
this case the purity of the system is determined by the statistics of
the injected field and becomes larger.

The non-degenerate OPO is often considered as a two-mode quantum
system, for which the pump can be treated as a classical quantity.  In
this point of view, it is described by a two-mode quantum state,
characterized by a density matrix which is the partial trace of
$\hat\rho_{\omega}$ over the pump mode. Its spectral purity is related
to the determinant of the two-mode covariance matrix. Such a "partial
purity" is displayed in figure 2. One observes that it is always
larger than the purity of the total system. This can seem surprising
that subsystem purity is larger that total purity, but it depends on
the presence or absence of quantum correlations between parts of the
system. For instance, when one only considers the differential mode,
by tracing over the pump and sum modes, one finds that its partial
purity is 1, and therefore that it is in a pure state. It is thereby
possible to extract from our initial mixed state a pure subsystem by
adequately eliminating the modes responsible for the "impurity" of the
total state

\subsection{Spectral purity for an asymmetrical injection}

In the previous subsection we have discussed the case of the
symmetrical injection of the OPO when both the idler and signal
modes with equal spectral widths $\kappa_i=\kappa_s$ are excited
equally by the external fields with amplitudes
$\sqrt{N^{in}_i}=\sqrt{N^{in}_s}$. Here we want to briefly discuss
the role of the asymmetry in the injection. The corresponding
derivations can be found in the appendix  of this paper, where the
physical situation with $N^{in}_i=N^{in}$ and $N^{in}_s=0$ is
considered. In this case, the mean output signal and idler fields
are no longer equal. In order to retrieve some symmetry we further
require that $\kappa_s=\kappa_i(1-\mu)$, so that the equality
$N_i=N_s\equiv N$ survives for the stationary situation.
Nevertheless the intracavity variances turn out to be different
and essentially differently dependent on $\mu$ than under the
symmetrical phase locking (see App~\ref{A}). Besides, the relation
between the variances of the output beam and the intracavity
variances turns out to be more complicated for the
$\pm$-variances:
\BY
2\langle\delta\hat
X^2_{\pm}\rangle_{\omega}=&&1+\frac{\kappa_i}{8N}[(\varepsilon^2_\pm)_\omega(1+\sqrt{1-\mu})^2+\nn\\
&&+
(\varepsilon^2_\mp)_\omega(1-\sqrt{1-\mu})^2],\\
2\langle\delta\hat
Y^2_{\pm}\rangle_{\omega}=&&1+\kappa_iN/2[(\varphi^2_\pm)_\omega(1+\sqrt{1-\mu})^2+\nn\\
&&+ (\varphi^2_\mp)_\omega(1-\sqrt{1-\mu})^2].
\EY

Fig. 3a shows the spectral dependence of the spectral purity in
the asymmetrical case for a small injection field parameter
$\mu=0.1$. One can see first that outside the spectral band
$\kappa$ the purity is close to 1, as in the symmetrical case. The
band where purity ${\Pi}_\omega\neq 1$, depends on the distance
to threshold and decreases when the pump power increases. Well
above threshold ($\mu_p\;>\;2$) it depends on $\mu$, and decreases
when the injecting field decreases. (see figure 3b)

However it is important to stress here that for the zero frequency
we can never neglect the influence of the injection field even for
very small values $\mu$. In the limit of small $\mu$ and $\mu_p>5$
we get ${\Pi}_{\omega=0}= 1/2$.

\subsection{Multi-frequency squeezing and entanglement
 in presence of small injection\L{V}}

We can use formulas (\ref{68})-(\ref{75}) to investigate the
squeezing and entanglement in the OPO in presence of small
injection. Even if this is not new in substance, we prove here that
the presence of a small injected field do not destroy the
non-classical features of the output fields.

Let us go back to the symmetrical case. Because
 \BY
&&4\langle\delta\hat X^2_{i,s}\rangle_{\omega}=\langle\delta\hat
X^2_{+}\rangle_{\omega}+\langle\delta\hat X^2_{-}\rangle_{\omega}
\EY
one has
\BY
&&4\langle\delta\hat
X^2_{i,s}\rangle_{\omega}=\nn\\
&&=1+\frac{1}{2}\;\frac{\kappa^2 }{\kappa^2(\mu_p-1+\mu/4)^2
+\omega^2}-\frac{1}{2}\;\frac{\kappa^2} {\kappa^2+\omega^2}.
\EY
Strongly above threshold $\mu_p\gg1$ the second term turns out to
be negligible and the maximum squeezing is reached
in the idler and signal waves.
\BY
&&4\langle\delta\hat X^2_{i,s}\rangle_{\omega=0}=\frac{1}{2}.
\EY
We can expect squeezing for $\mu_p>2$.

In contrast with the amplitude squeezing the phase squeezing is
found only in the pump mode:
\BY
&&4\langle\delta\hat
Y^2_{p}\rangle_{\omega}=1-\frac{2\kappa^2(\mu_p-1)
}{\kappa^2\mu_p^2 +\omega^2}.
\EY
On the zero frequency
\BY
&&4\langle\delta\hat Y^2_{p}\rangle_{\omega=0}=\frac{(\mu_p-1)^2+1
}{\mu_p^2 }.
\EY
It is not difficult to see that maximum squeezing
$4\langle\delta\hat Y^2_{p}\rangle_{\omega=0}=1/2$ is reached as
$\mu_p=2$.

In order to evaluate an entanglement of the idler and signal
waves, we use the Duan criterium
\BY
&&2\langle\delta\hat X_-^2\rangle,\;2\langle\delta\hat
Y_+^2\rangle<1.
\EY
Taking into account (\ref{68}) and (\ref{71}) one can find that
this criterium is carried out independently of pump. Even if,
strongly above threshold  ($\mu_p\gg1 $) $2\langle\delta\hat
Y^2_{+\omega=0}\rangle\to1.$

\section{Stationary Glauber quasi-probability distribution \L{IV}}
In the previous section we considered  the properties of the
ND-OPO from the point of view of its spectral components, by
solving non stationary equations in Fourier space. Stationary
master equations also contain important pieces of information about
the quantum system. We devote this paragraph to the
determination of the Glauber quasi-probability distribution of the
present three-mode system in the limit of small amplitude and
phase fluctuations which is relevant for the case of the injected
ND-OPO. It is obtained by putting all time derivatives to zero in
Eqs~(\ref{23.})-(\ref{26.}). We will discuss in the two following
subsections the solutions of the amplitude and phase equations
respectively.

\subsection{Amplitude quasi-probability distribution}
The stationary amplitude quasi-probability
$P(\varepsilon_p,\varepsilon_i,\varepsilon_s)$ can be factorized
in the form
\BY
&&P(\varepsilon_p,\varepsilon_i,\varepsilon_s)=P(\varepsilon_p,\varepsilon_+)P(\varepsilon_-),\L{96}
\EY
where each of the factors obeys its own equations:
 \BY
&&\(\kappa(1-\mu/2)\frac{\partial}{\partial\varepsilon_-}\varepsilon_-
-\kappa N(1-\mu)
\frac{\partial^2}{\partial\varepsilon_-^2}\)P(\varepsilon_-)=0,\L{97}\nn\\
&&\\
&&\left(\frac{1}{2}\frac{\partial}{\partial\varepsilon_p}(\kappa_p\varepsilon_p
+\kappa(1-\mu)\varepsilon_+)
+\frac{\partial}{\partial\varepsilon_+}(\kappa\mu/2\;\varepsilon_+\right.
-\nn\\
&&-\left.\kappa_p(\mu_p-1)\varepsilon_p) + \kappa
N(1-\mu)\frac{\partial^2}{\partial\varepsilon_+^2}\right)P(\varepsilon_p,\varepsilon_+)=0.\L{98}\nn\\
\EY

Let us first consider the second equation. The solution has the
Gaussian form :
 \BY
&&
P(\varepsilon_p,\varepsilon_+)=\frac{1}{2\pi\sqrt{D_pD_+}}\exp\(-\frac{(\varepsilon_p-a\varepsilon_+)^2}
{2D_p} -\frac{\varepsilon_+^2}{2D_+}\).\L{103}\nn\\
&&
\EY
The unknown parameters $D_p,D_+$
and $a$ are coupled with the variances by means of the following
relations :
\BY
D_+=\o{\varepsilon_+^2},\qquad D_p=\o{\varepsilon_p^2}-
\frac{\o{\varepsilon_+\varepsilon_p}^2}{\o{\varepsilon_+^2}},\qquad
a= \frac{\o{\varepsilon_+\varepsilon_p}}{\o{\varepsilon_+^2}}.
\EY
As the equation (\ref{98}) provides us with a possibility to find the
variances in the explicit form, then taking into account the previous
inequalities we can calculate the parameters of the distribution
$D_p,D_+$ and $a$. From eq. (\ref{98}) one can then obtain the
algebraic system of equation for the variances in the form :
\BY
&&-\kappa_p\;\o{\varepsilon_p^2}-\kappa(1-\mu)\;\o{\varepsilon_p\varepsilon_+}=0,\\
&&-\frac{1}{2}(\kappa_p+\kappa\mu)\;\o{\varepsilon_p\varepsilon_+}-\frac{1}{2}\kappa(1-\mu)\;\o{\varepsilon_+^2}+
\kappa_p(\mu_p-1)\;\o{\varepsilon_p^2}=0,\nn\\&&\\
&&-\kappa\mu\;\o{\varepsilon_+^2}+
2\kappa_p(\mu_p-1)\;\o{\varepsilon_p\varepsilon_+}+2\kappa
N(1-\mu)=0,
\EY
Solving this system, one can get  the variances :
\BY
&&\o{\varepsilon_+^2}= N\frac{1}{\mu/2+\mu_p-1}\(1+
\frac{2\kappa}{\kappa_p+\kappa\mu}(\mu_p-1)\),\\
&&\o{\varepsilon_+\varepsilon_p}=
-N_p\;\frac{1}{\mu/2+\mu_p-1}\;\frac{\kappa_p}{\kappa_p+\kappa\mu}(\mu_p-1),\\
&& \o{\varepsilon_p^2}=
N_p\frac{1}{\mu/2+\mu_p-1}\;\frac{\kappa}{\kappa_p+\kappa\mu}(\mu_p-1).
\EY
In the limit $\kappa,\kappa(\mu_p-1)\ll\kappa_p$ and $\mu\ll1$
these variances read
\BY
&&\o{\varepsilon_+^2}= N\;\frac{1}{\mu/2+\mu_p-1},\nn\\
&&
\o{\varepsilon_p^2}=N_p\;\frac{\mu_p-1}{\mu/2+\mu_p-1}\;\frac{\kappa}{\kappa_p},\\
&&
\o{\varepsilon_+\varepsilon}_p=-N_p\;\frac{\mu_p-1}{\mu/2+\mu_p-1}.\nn
\EY

The knowledge of the variances provides us with a possibility to
find the parameters of the Gaussian quasi-probability:
\BY
&&D_+= N\;\frac{1}{\mu/2+\mu_p-1},\nn\\
&&
D_p=N_p\;\frac{\mu_p-1}{\mu/2+\mu_p-1}\;\frac{\mu\kappa}{\kappa_p},\\
&& a=-\frac{\kappa}{\kappa_p}.\nn
\EY
Let us discuss now the quasi-probability $P(\varepsilon_-)$ that
is a solution of equation (\ref{97}). This equation has a
normalized solution only in the form of the distribution. This is
connected with the minus sign in front of the second derivative
with respect to $\varepsilon_-$. On the one hand this minus sign
informs us about the quantum effects in the generation and on the
other hand it makes impossible the derivation of the solution as a
well-behaved function. The direct result of this situation is the
variance $\o{\varepsilon_-^{2}}$ turns out to be negative
\BY
&&\o{\varepsilon_-^2}=-N
\EY
although the following relation
\BY
&&\o{\varepsilon_-^{2k}}=(2k-1)!!\;\o{\varepsilon_-^{2}}^k,
\EY
typical for the Gaussian distribution, survives. Knowing the
variances allows us to derive the P-function as the formal series
\cite{Sudarshan,Glauber}:
\BY
&&P(\varepsilon_-)=\sum\limits_{k=0}^\infty\frac{1}{k!}\(-\frac{N}{2}\)^k
\frac{d^{2k}}{d\varepsilon_-^{2k }}\;\delta(\varepsilon_-).\L{115}
\EY
It is also possible to use another formal equivalent expression:
\BY
&&P(\varepsilon_-)
=exp\(-\frac{N}{2}\frac{d^{2}}{d\varepsilon_-^{2
}}\)\delta(\varepsilon_-) \L{dist}
\EY
From these expressions, one can derive Fano parameters for each of
the intracavity modes:
\BY
&&F_{i,s}= \frac{1}{4}\;\frac{2-\mu_p}{\mu/2+\mu_p-1}+1,\qquad
F_{p}=1.
\EY

\subsection{Phase quasi-probability distribution}
As for the amplitude, the quasi-probability distribution of the
three phases can be factorized in the form:
\BY
&&P(\varphi_p,\varphi_i,\varphi_s)=P(\varphi_p,\varphi_+)P(\varphi_-).
\EY
The corresponding master equation read (\ref{23.})-(\ref{26.}):
 \BY
&&\left(\frac{\kappa_p}{2}\frac{\partial}{\partial\varphi_p}
(\varphi_p+(\mu_p-1)\varphi_+)+\frac{\partial}{\partial\varphi_+}
(\kappa(1-\mu/2)\varphi_+-\right.\nn\\
&&\left.-\kappa(1-\mu)\varphi_p)- \frac{\kappa}{4N}(1-\mu)\;
\frac{\partial^2}{\partial\varphi_+^2}\right)P_(\varphi_p,\varphi_+),\L{119}\\
&&\(\frac{\mu\kappa}{2}\frac{\partial}{\partial\varphi_-}
\varphi_-+
\frac{\kappa}{4N}(1-\mu)\;\frac{\partial^2}{\partial\varphi_-^2}\)P(\varphi_-).\L{120}
\EY
Again the minus sign in front of the last term means quantum phase
features in the field. A formal solution of the equation can be
presented in the form of the distribution:
 \BY
&&P(\varphi_p,\varphi_+)=\sum\limits_{m,n=0}^\infty\frac{(-1)^{m+n}}{m!\;n!}
M_{mn}\frac{d^{m}}{d\varphi_p^{m}}
\frac{d^{n}}{d\varphi_+^{n}}\;\delta(\varphi_p)\;\delta(\varphi_+),\nn\\
&& M_{mn}=\o {\varphi_p^m\;\varphi_+^n} \L{121}
 \EY
It is possible to demonstrate that the momenta $M_{mn}$ are
different from zero only as $m+n$ is the even number. Then, with
help of equation (\ref{119}), we are able to derive a recurrence
relation connecting different non-zero even momenta with each
other:
 \BY
&&-(m\;\kappa_p/2+n\;\kappa)M_{mn}-m\;(\mu_p-1)\kappa_p/2\;M_{m-1n+1}
 +\nn\\
&&+n\kappa M_{m+1n-1}-n(n-1)\kappa/(4N)\;M_{mn-2}=0.\nn\\
\EY
In particular, as $\kappa,(\mu_p-1)\kappa\ll\kappa_p$
\BY
&&\o{\varphi_p^2}=-\frac{1}{4N_p}\;\frac{\mu_p-1}{\mu_p}
\;\frac{\kappa}{\kappa_p},\nn\\
&&\o{\varphi_+^2}=-\frac{1}{4N}\;\frac{1}{\mu_p},\\
&&\o{\varphi_p\;\varphi_+}=\frac{1}{4N_p}\;\frac{1}{\mu_p}\;\frac{\kappa}{\kappa_p}.\nn
\EY
As for Eq~(\ref{120}) it describes a stationary Gaussian process
and its solution is well-known:
 \BY
&&P(\varphi_-)=\frac{1}{\sqrt{\pi/(\mu N)}}\;
\exp\(-\frac{\varphi_-^2}{1/(\mu N)}\)\L{124}
\EY

\subsection{Stationary output quasi-probability distribution and purity}
In the previous section, the output beams of the ND-OPO were
presented as a set of mutually coupled field oscillators with
different frequencies. In contrast, the P-function that we have
just derived describes the system in its stationary state, but
only for the three intracavity oscillators. Unfortunately, to the
best of our knowledge, there is no simple relation between the
intracavity P-function and some kind of quasi-probability
distribution describing the system of the three OPO-modes when
they have escaped the cavity. However it is possible to maintain a
single-oscillator description of the three modes outside the
cavity in a very specific case, as it has been explained by one of
us Ref~\cite{Gol}. The distribution function then only concerns
output fields contained in "thin layers" of the propagation axis.
The thickness of each of these layers must be much bigger than the
wavelength, but much less than the cavity correlation length
$l_0=c\tau$, where $\tau$ the correlation time of the intracavity
field. Then it is formally possible to introduce photon operators
$\hat a_m$ and $\hat a_m^\dag$ acting in m-th layer such as
$\[\hat a_m,\hat a_n^\dag\]=\delta_{mn}$. This means that the
field outside the cavity can be presented as a set of spatially
located oscillators and that the propagation of the light in free
space can be presented as a transfer of an excitation from one of
the oscillators to the nearest one along the beam. These
observables correspond to measurements performed on the output
beams that integrate the photocurrent fluctuations over time
scales much smaller than $\tau$. One can then define a
quasi-probability distribution $P_{out}$ for such output
oscillators, which is related to the already calculated
intracavity P-function by means of
\BY
&&P_{out}(\alpha_{p,out},\alpha_{i,out},\alpha_{s,out})=\nn\\
&& =\frac{1}{T_pT_iT_s}\;
P\(\frac{\alpha_{p,out}}{\sqrt{T_p}},\frac{\alpha_{i,out}}{\sqrt{T_i}},\frac{\alpha_{s,out}}{\sqrt{T_s}
}\).\L{29}
\EY
$\sqrt{T_i}=\sqrt{T_s}\equiv\sqrt{T}$ being the amplitude
transmission coefficients of the coupling mirror for the signal
and idler modes; $\sqrt{T_p}$ the transmission coefficient for the
pump mode; $\alpha_{p,out},\alpha_{i,out}\alpha_{s,out}$ are the
corresponding complex amplitudes, which are the eigenvalues of the
annihilation photon operator.

Knowing $P_{out}$ of the whole system outside the cavity, at least
in the restricted meaning described in the previous paragraph, it
is now possible to calculate the purity of the stationary state of
this system. It is the product of four factors:
\BY
&&{\Pi}_{st}={\Pi}_{1}\times{\Pi}_{2}\times{\Pi}_{3}\times{\Pi}_{4},
\EY
where the amplitude factors are
\BY
{\Pi}_{1}&&=\int\int d\varepsilon_p d\varepsilon_+
\;P_{out}(\varepsilon_p,\varepsilon_+) \int d\varepsilon_p^\prime
 d\varepsilon_+^\prime\;
P_{out}(\varepsilon_p^\prime,\varepsilon_+^\prime)\times\nn\\
&&\times\exp\(
-\frac{(\varepsilon_p-\varepsilon_p^\prime)^2}{4N_pT_p}\)\exp\(
-\frac{(\varepsilon_+-\varepsilon_+^\prime)^2}{8NT}\),\L{110}\\
{\Pi}_{2}&&=\int d\varepsilon_- P_{out}(\varepsilon_-)\int
 d\varepsilon_-^\prime\;P_{out}(\varepsilon_-^\prime)\times\nn\\
&&\times\exp\(
-\frac{(\varepsilon_--\varepsilon_-^\prime)^2}{8NT}\) ,\L{}
\EY
and the phase ones are:
\BY
{\Pi}_{3}&&=\!\!\iint d\varphi_pd\varphi_+
P_{out}(\varphi_p,\varphi_+)\!\iint d\varphi_p^\prime
d\varphi_+^\prime
P_{out}(\varphi_+^\prime,\varphi_p^\prime)\!\times\nn\\
&&\times
\exp\(-\frac{(\varphi_p-\varphi_p^\prime)^2}{1/(N_pT_p)}\)
\exp\(-\frac{(\varphi_+-\varphi_+^\prime)^2}{2/(NT)}\),\\
{\Pi}_{4}&&=\int d\varphi_- \;P_{out}(\varphi_-)\int
d\varphi_-^\prime\;
P_{out}(\varphi_-^\prime)\times\nn\\
&&\times\exp\(-\frac{(\varphi_--\varphi_-^\prime)^2}{2/(NT)}\).\L{113}
\EY
Substituting Eqs.~(\ref{103}), (\ref{115}), (\ref{121}), and
(\ref{124}) and considering the limit
$\kappa,\;(\mu_p-1)\kappa\ll\kappa_p$ we get
\BY
&&{\Pi}_{1}=\frac{1}{(1+\nu \;T/\mu)^{1/2}},\nn\\
&&{\Pi}_{2}={\Pi}_{3}=1,\\
&&{\Pi}_{4}=\frac{1}{(1+T/\mu)^{1/2}},\nn
\EY
where
\BY
&&
\nu=\frac{\mu(\mu_p-1/2)}{\mu/2+\mu_p-1},\nn\\
\EY
If the output mirror of the cavity is highly reflecting, so that
that $T\ll\mu$, then
\BY
&&{\Pi}_{st}=1.
\EY
A similar calculation performed with the intracavity P-function with
the same parameters gives ${\Pi}_{1}={\Pi}_{2}={\Pi}_{3}=1$, and
${\Pi}_{4}=\sqrt{\mu}$, so that the purity of the intracavity state,
${\Pi}=\mu^{1/2}$ is much smaller than 1.

Field fluctuations integrated over short time intervals depend
only on the high frequency fluctuations of the output fields. The
property that ${\Pi}_{st} = 1$ is therefore certainly
connected to the already noticed feature that ${\Pi}_\omega=
1$ for noise frequencies larger than the cavity bandwidth.

\section{Stationary photon number probability  \L{V}}
Another important way of characterizing the quantum state produced
by the TROPO is to determine the full photon number probability
distribution in each of the three beams exiting the system. This
quantity is important to know, for example when one wants to
determine the quantum state which is produced by a conditional
measurement performed on the intensity of one of the output beams.
This is what we will do in this section. More precisely, we will
determine the probability for n photons to cross the cross section
of the output beam during a given time $\tau$. Let us stress that
this function is essentially different than the already calculated
stationary probability for the photon number because it includes
the integration time $\tau$.

Let us first calculate the photon number probability inside the
cavity. The joint probability $C_{in}(n_i, n_s)$ to find $n_i$
photons in the idler intracavity mode and simultaneously $n_s$
photons in the signal intracavity mode is defined as

\BY
 &&
C_{in}(n_i,n_s)=\sum\limits_{n_p=0,1,\ldots}  \langle
n_pn_in_s|\hat\rho|n_pn_sn_i\rangle, \L{78}
 \EY
where $\hat\rho$ is the stationary three-mode intracavity density
matrix. By using the Glauber diagonal representation (7) in (78),
we have
 \BY
C_{in}(n_i,n_s)=\frac{1}{4
}\int\limits_0^\infty\!\!\int\limits_0^\infty&&
du_i\;du_sP_{red.}(u_i,u_s)\times\nn\\
&&\times e^{-u_i}\frac{u_i^{n_i}}{n_i!}\;e^{-u_s}\frac{u_s^{n_s}
}{n_s!}, \L{135}
 \EY
where $P_{red.}$ is a "reduced" Glauber photon quasi-probability
given by
\BY
P_{red.}(u_i,u_s)=\!\!\iint d^2\alpha_p\!\iint
\!d\varphi_id\varphi_s P(\alpha_p,\alpha_i,\alpha_s),
\EY
where $\alpha_l = \sqrt{u_l} \;\exp(i\varphi_l) (l = s, i)$. From
the previous section
\BY
&&P_{red.}(u_i,u_s)=P(\varepsilon_-)\int d\varepsilon_p
P(\varepsilon_p,\varepsilon_+)=\\
&&=\frac{1}{\sqrt{2\pi D_+}}\exp\( -\frac{\varepsilon_+^2}{2D_+}\)
\exp\(-\frac{N}{2}\frac{d^2}{d\varepsilon_-^2}\)\delta(\varepsilon_-)\nn
\EY
Substituting this into (\ref{135}) one can get after the
corresponding integrating one can obtain that
 \BY
 C_{in}(n_i,n_s)&&=1/\sqrt{2\pi \lambda N}
\;\exp\(-\frac{(n_+-2N)^2}{2\lambda N}\)\times\nn\\
&&\times1/\sqrt{2\pi N }\exp\(-\frac{n_-^2}{2 N}\),\L{137}
\EY
where
\BY
&& \lambda=
2+\frac{1}{\mu/4+\mu_p-1}\\
&&(\kappa,\;(\mu_p-1)\kappa\ll\kappa_p).\nn
\EY
This function describes the probabilities to find $n_i$ photons in
the idler mode and $n_s$ photons in the signal mode inside the
cavity in the stationary regime. Although knowing the intracavity
probability is not enough for determining the probability for the
output travelling fields, it will helps us to guess the general
form of the wished function. We will assume that the outside
probability for output fields has the same Gaussian form as the
probability for the intracavity fields, but with its own
parameters depending, in particular, on observation time $\tau$.
We will therefore write it as:

 \BY
 C_{out}(n_i,n_s)&&=1/\sqrt{2\pi d_+}
\;\exp\(-\frac{(n_+-2N_{out})^2}{2d_+}\)\times\nn\\
&&\times1/\sqrt{2d_- }\exp\(-\frac{n_-^2}{2d_-}\).\L{137}
\EY
where the parameters $N_{out},\;d_\pm$ depend on $\tau$. The way
to calculate the parameters will be demonstrated on the simpler
photon probability of a single mode.

By integrating (\ref{137}) over one of the variances
$\varepsilon_i$ or $\varepsilon_s$ we  obtain the single-mode
photon number probability for both the idler and signal modes:
\BY
&& C_{in}(n_{l})=\frac{1}{\sqrt{2\pi
NF_{in}}}\;\exp\(-\frac{(n_{l}-N)^2}{2NF_{in}}\),\\
&& F_{in}=\frac{1}{4}(1+\lambda),\qquad l=i,s,\nn\L{141}
\EY
and correspondingly for the output beam:
\BY
&& C_{out}(n_{l})=\frac{1}{\sqrt{2\pi
F_{out}}}\;\exp\(-\frac{(n_{l}-N_{out})^2}{2NF_{out}}\).\L{142}
\EY
In the last probability $n_l$ are eigen-numbers of the operator
\BY
&&\hat n_l(t)=\int\limits_{t-\tau/2}^{t+\tau/2} \hat
A_l^{\dag}(t^\prime)\;\hat A_l(t^\prime)\;dt^\prime,
\EY
where the operators $\hat A_l$ and $\hat A_l^\dag$ describe light
propagating in free space (see (\ref{k})). Then
\BY
&&N_{out}=\int\limits_{t-\tau/2}^{t+\tau/2} \langle\hat
A_{l}^{\dag}(t^\prime)\;\hat
A_{l}(t^\prime)\rangle\;dt^\prime=\kappa\tau N.\L{}
\EY
In order to determine the Fano-factor $F_{out}$, we have to
calculate the variance
\BY
 && \langle \hat n_{l}^2\rangle-\langle \hat
n_{l}\rangle^2=N_{out}F_{out}.
\EY
For the stationary flux
\BY
&&N_{out}F_{out}=\\
&&=\int\limits_{-\tau/2}^{\tau/2}\!\!\int\limits_{-\tau/2}^{\tau/2}dt_1dt_2
\langle\hat A_{l}^\dag(t_1)\hat A_{l}(t_1)\hat A_{l}^\dag(t_2)\hat
A_{l}(t_2)\rangle- N_{out}^2.\nn
\EY
Following to standard procedure, one obtains
\BY
&&F_{out}=1+\kappa/N\int
d\omega\;(\varepsilon_l^2)_{\omega}\;\delta_\tau(\omega),
\EY
where the integral contains a product of two functions. One of
them is
\BY
&&\kappa/N\;(\varepsilon_l^2)_{\omega}=\frac{
1}{2}\(\frac{\kappa^2}{(\mu/2+\mu_p-1)^2\kappa^2+\omega^2}-\;\frac{\kappa^2}{\kappa^2+\omega^2}\).\nn\\
&&
\EY
It is the spectral density of the amplitude noise in the selected
l-wave that can be obtained from (\ref{41.}) and (\ref{44.}). The
other is
\BY
&&
\delta_\tau(\omega)=\frac{1}{\pi}\;\frac{\sin^2\omega\tau/2}{\tau\omega^2/2}.
\EY
For long enough integration times $\tau$, $\delta_\tau(\omega)$
gets close to a $\delta$-function when $\tau\to\infty$, so that
$F_{out}$ is given by

\BY
&&F_{out}=\frac{1}{2}\;\frac{\mu_p}{\mu/2+\mu_p-1}.
\EY
Near threshold $F_{out}=\mu^{-1}\gg1$, so that the photon
statistics of the signal or idler field turns out to be
super-Poissonian. In contrast, strongly above threshold
$F_{out}=1/2$, and the photon statistics of the signal or idler
field is now sub-Poissonian, as it has been already noticed (ref
?) and experimentally checked (Kasai) .

The joint signal-idler photon probability distribution can be
derived exactly in the same way, knowing the two-mode variances
$d_\pm$ (\ref{137}):
\BY
&&d_{\pm}=2N_{out} \(1+\kappa/(2N)\int
d\omega\;(\varepsilon_\pm^2)_{\omega}\;\delta_\tau(\omega)\)\L{132.}.
\EY
As $(\mu_p-1)\kappa,\kappa\ll\kappa_p$
\BY
&&\kappa/(2N)\;(\varepsilon_+^2)_{\omega}=
\frac{\kappa^2}{(\mu/4+\mu_p-1)^2\kappa^2+\omega^2},\nn\\
&&\kappa/(2N)\;(\varepsilon_-^2)_{\omega}=-
\frac{\kappa^2}{\kappa^2+\omega^2}
\EY
Inserting these expressions to formula (\ref{132.}), one obtains
the full joint photon number probability distribution
$C_{out}(n_i,n_s)$ of the output signal and idler beams for any
observation time $\tau$.

If we choose the short observation time $\tau$ such as\\
$\kappa\tau,\; \kappa\tau(\mu/4+\mu_p-1)\ll 1$ the variances read:
\BY
d_\pm=2 N_{out}.
\EY
This means that under the observation for only short time the
photon statistics turn out to be Poissonian.

On the other hand, for $\kappa\tau,\; \kappa\tau(\mu/4+\mu_p-1)\gg
1$,
\BY
d_+=\mu_p/(\mu/2+\mu_p-1),\quad d_-=0.
\EY
We have a possibility to derive the wished photon number
probability in form:
 \BY
 &&C_{out}(n_i,n_s)=\\
&&=\frac{1}{\sqrt{2\pi d_+}}
\exp\(-\frac{(2n_i+-2N_{out})^2}{2d_+}\)\delta(n_i-n_s).\nn\L{}
\EY
So if we count photons during long time intervals, the photon
numbers in both waves turn out to be the same. This the form of
the joint probability distribution that was guessed in \cite{Laurat2004},
and used to calculate the state of the signal mode produced by a
conditional measurement performed on the idler mode.

\section{Conclusion}

One of our aims here was to investigate the quantum state purity
of the TROPO radiation. There is a widely-distributed opinion that
the purity of the output emission state must be equal to one. The
reason of that is connected with a representation about the OPO as
system that, in the unitary process, converts  the pure state of
the input light (the pump wave is in the coherent state and the
other modes are in the vacuum states or the pump, idler, signal
waves are in the coherent states and again the other modes are in
the vacuum states) into the same pure state of the output light.
This  conclusion absolutely correct for the output field as whole
nevertheless  does not concern the states for the selected
frequencies, and just it is  an object of the investigation in
experiments and theories. For example, the covariance matrix is
constructed as a combination of the quadrature variances for the
selected frequency. The purity for this oscillator turns out to be
uncertain and, in principle, can accept any meanings.

We have considered the  purity and been convinced that, strictly
speaking,  the purity for the state with selected frequency is not
one (especially  near zero frequencies). However for the
symmetrical synchronization the purity turns out to be close to
one, provided the pump power  is strongly above threshold. At the
same time under the asymmetrical synchronization the purity
becomes $1/2$ even if $\mu_p\gg1$. So we can conclude that,
generally speaking, the purity for the oscillator on the selected
frequency is not one and essentially depends on the power of the
synchronizing field and the asymmetry of the synchronization too.

We needed to introduce the synchronization into our the OPO system
to depress the  diffusion of the differential phase between signal
and idler waves. Usually, it is suggested that the synchronization
especially by the weak external field does not insert any
essential distortions into a statistical pattern of the TROPO
except a phase diffusion depression. However, strictly speaking,
this is quite not obvious and one of our aims here was to
investigate just this side of the OPO generation. As was mentioned
already the purity can be essentially dependent on the
synchronization.

Besides although we introduce the synchronization to depress the
phase diffusion, nevertheless  not only the phase fluctuations are
stabilized under an influence of the external field but the
amplitude ones too. We have found that this phenomenon turns out
to be essential near threshold as $\mu_p-1<\mu$. One can see that
all amplitude variances on the zero frequency are proportional to
$1/\mu^2$ (not to infinity as under the phase diffusion). This
apiaries  as well as in the spectral and stationary variances.

\section{Acknowledgements}

This work was performed within the French-Russian cooperation
program "Lasers and Advanced Optical Information Technologies"
with financial support from the following organizations: INTAS
(Grant No. 7904),YS-INTAS (Grant No. 6078), RFBR (Grant No.
05-02-19646), and Ministry of education and science of RF (Grant
No. RNP 2.1.1.362).

\appendix

\section{Case of an asymmetrical injection\L{A}}
\subsection{Master equations}
In the main part of the article, we discussed in detail how the
phase locking phenomenon produced by a symmetrical injection on
the idler and signal modes by an external coherent light acts on
the statistical properties of the OPO radiation. In this appendix
we consider the case of an asymmetrical injection, more precisely
when the idler mode is injected by a coherent field with amplitude
$\sqrt{N_i^{in}}=\sqrt{N^{in}}$ and the signal mode by the vacuum
state $\sqrt{N_s^{in}}=0$. Let us first rewrite the master
equation in the form:
 \BY
&&\frac{\partial P(\alpha_p,\alpha_i,\alpha_s,t)}{\partial
t}=\frac{\kappa_i}{2}\frac{\partial
}{\partial\alpha_s}\alpha_sP+\frac{\kappa_s}{2}\frac{\partial
}{\partial\alpha_i}(\alpha_i-\sqrt{N^{in}})P
+\nn\\
&&+\frac{\kappa_p}{2}\frac{\partial
}{\partial\alpha_p}(\alpha_p-\sqrt{N_p^{in}}) P
+g\left(\alpha_i\alpha_s\frac{\partial
P}{\partial\alpha_p}-\alpha_p\alpha_s^\ast\frac{\partial
P}{\partial\alpha_i}-\right.\nn\\
&&\left.-\alpha_p\alpha_i^\ast\frac{\partial
P}{\partial\alpha_s}\right)+g\alpha_p\frac{\partial^2
P}{\partial\alpha_i\partial\alpha_s}+c.c.\L{8}
\EY
The corresponding classical equations read
\BY
&&\dot\alpha_p=-\frac{\kappa_p}{2}\;(\alpha_p-\sqrt{N_p^{in}})-g\alpha_i\alpha_s\\
&&\dot\alpha_i=-\frac{\kappa_i}{2}\;(\alpha_i-\sqrt{N^{in}})+g\alpha_p\alpha_s^\ast\\
&&\dot\alpha_s=-\frac{\kappa_s}{2}\;\alpha_s+g\alpha_p\alpha_i^\ast
\EY
Putting
\BY
&&\kappa_i=\kappa_s(1-\mu)
\EY
one can obtain the stationary solutions of the classical equations
in the form:
 \BY
&&\alpha_p=\sqrt{N_p},\qquad \alpha_{i,s}=\sqrt{N},
\EY
where
\BY
&& N_p=\frac{\kappa_s^2}{4g^2},\qquad
N=(\mu_p-1)\;\frac{\kappa_s\kappa_p}{4g^2},
\EY
and
 \BY
&& \kappa_s N=(1-\mu)\kappa_i N=(\mu_p-1)\kappa_p N_p.
\EY
In the limit of small amplitude and phase fluctuations it is
possible to factorize the solutions in its phase and amplitude
factors:
\BY
&&P(\alpha_p,\alpha_i,\alpha_s,t)=P(\varepsilon_p,\varepsilon_i,\varepsilon_s,t)
P(\varphi_p,\varphi_i,\varphi_s,t).
\EY
We now introduce to usual sum and difference notations
\BY
&&\varepsilon_\pm=\varepsilon_i\pm\varepsilon_s,
\qquad\varphi_\pm=\varphi_i\pm\varphi_s
\EY
and convert equation (\ref{8}) into two equations:
 \BY
&&\frac{\partial
P(\varepsilon_p,\varepsilon_+,\varepsilon_-,t)}{\partial
t}=\left(\frac{1}{2}\frac{\partial}{\partial\varepsilon_p}(\kappa_p\varepsilon_p
+\kappa_s\varepsilon_+)
+\right.\nn\\
&&+\frac{\partial}{\partial\varepsilon_+}\left(\frac{\mu\kappa_i}{4}\varepsilon_+
-\frac{(\mu_p-1)\kappa_p}{\sqrt{1-\mu}}\varepsilon_p\right) +
\kappa_s
N\frac{\partial^2}{\partial\varepsilon_+^2}+\nn\\
&&\left.+\kappa_i\;(1-\frac{3\mu}{4})\frac{\partial}{\partial\varepsilon_-}\varepsilon_-
-\kappa_s N
\frac{\partial^2}{\partial\varepsilon_-^2}\right)P(\varepsilon_p,\varepsilon_+,\varepsilon_-,t)+\nn\\
&&+\mu\(\frac{\partial P}{\partial
t}\)_1,\\
&&\frac{\partial P(\varphi_p,\varphi_+,\varphi_-,t)}{\partial
t}=\left(\frac{\kappa_p}{2}\frac{\partial}{\partial\varphi_p}
(\varphi_p+\frac{(\mu_p-1)}{\sqrt{1-\mu}}\;\varphi_+)+\right.\nn\\
&&+\frac{\partial}{\partial\varphi_+}
(\kappa_i(1-\frac{3\mu}{4})\;\varphi_+-\kappa_s\varphi_p)
-\nn\\
&&-\frac{\kappa_s}{4N}\;
\frac{\partial^2}{\partial\varphi_+^2}+\frac{\mu\kappa_i}{4}\frac{\partial}{\partial\varphi_-}
\varphi_-+\nn\\
&&
+\left.\frac{\kappa_s}{4N}\;\frac{\partial^2}{\partial\varphi_-^2}\right)P(\varphi_p,\varphi_+,\varphi_-,t)
+\mu\(\frac{\partial P}{\partial t}\)_1.\L{26}
\EY
Here the selected terms on the right of the equations have the
following form
 \BY
\lefteqn{\(\frac{\partial P}{\partial
t}\)_1=\frac{\kappa_i}{4}\(\frac{\partial}{\partial\varepsilon_+}\varepsilon_-
+\frac{\partial}{\partial\varepsilon_-}\varepsilon_+\)
P(\varepsilon_p,\varepsilon_+,\varepsilon_-,t),}\\
&\(\ds\frac{\partial P}{\partial
t}\)_1=\ds\frac{\kappa_i}{4}\(\ds\frac{\partial}{\partial\varphi_+}
\varphi_-+\ds\frac{\partial}{\partial\varphi_-}
\varphi_+\)P(\varphi_p,\varphi_+,\varphi_-,t).\nn\\
\EY
One can see that their role is essential when the synchronizing
parameter $\mu$ is big. Then some additional mixing the variables
$\varepsilon_l$ takes place. To avoid it we  require that
$\mu\ll1$. A similar remark concerns the variables $\varphi_l$. In
our limit we neglect these terms, so that the Glauber
quasi-probability is factorized in the form
\BY
&&P(\alpha_p,\alpha_i,\alpha_s,t)=P(\varepsilon_p,\varepsilon_+,t)
P(\varepsilon_-,t)P(\varphi_p,\varphi_+,t) P(\varphi_-,t)\nn\\
\EY
and correspondingly the equations are decoupled as
 \BY
&&\frac{\partial P(\varepsilon_p,\varepsilon_+,t)}{\partial
t}=\left(\frac{1}{2}\frac{\partial}{\partial\varepsilon_p}(\kappa_p\varepsilon_p
+\kappa_s\varepsilon_+)
+\right.\nn\\
&&+\frac{\partial}{\partial\varepsilon_+}\left(\frac{\mu\kappa_i}{4}\varepsilon_+
-\frac{(\mu_p-1)\kappa_p}{\sqrt{1-\mu}}\varepsilon_p\right)
+ \nn\\
&&+\left.\kappa_s
N\frac{\partial^2}{\partial\varepsilon_+^2}\right)P(\varepsilon_p,\varepsilon_+,t),\L{23}\\
&&\frac{\partial P(\varepsilon_-,t)}{\partial
t}=\(\kappa_i(1-3\mu/4)\frac{\partial}{\partial\varepsilon_-}\varepsilon_-
-\kappa_s N
\frac{\partial^2}{\partial\varepsilon_-^2}\)P(\varepsilon_-,t),\nn\\
&&\\
&&\frac{\partial P(\varphi_p,\varphi_+,t)}{\partial
t}=\left(\frac{\kappa_p}{2}\frac{\partial}{\partial\varphi_p}
(\varphi_p+\frac{(\mu_p-1)}{\sqrt{1-\mu}}\varphi_+)+\right.\nn\\
&&\left.+\frac{\partial}{\partial\varphi_+}
(\kappa_i(1-3\mu/4)\varphi_+-\kappa_s\varphi_p)-
\frac{\kappa_s}{4N}\;
\frac{\partial^2}{\partial\varphi_+^2}\right)P_(\varphi_p,\varphi_+,t),\L{}\nn\\
&&\\
&&\frac{\partial P(\varphi_-,t)}{\partial
t}=\(\frac{\mu\kappa_i}{4}\frac{\partial}{\partial\varphi_-}
\varphi_-+
\frac{\kappa_s}{4N}\;\frac{\partial^2}{\partial\varphi_-^2}\)P(\varphi_-,t).\L{26}
\EY

In order to analyze the time dependent (spectral) correlation
functions, we use the Langevin equations of the system, which are
have the following form:
\BY
&&\dot\varepsilon_p=-\kappa_p/2\;\varepsilon_p-\kappa_s/2\;\varepsilon_+,\\
&&\dot\varepsilon_+=-\kappa_i\mu/4\;\varepsilon_++\kappa_p(\mu_p-1)/\sqrt{1-\mu}\;\varepsilon_p+f_+(t),\nn\\
&&\\
&&\dot\varepsilon_-=-\kappa_i(1-3\mu/4)\;\varepsilon_-+f_-(t),\\
&&\dot\varphi_p=-\kappa_p/2\;\varphi_p-\kappa_p(\mu_p-1)/(2\sqrt{1-\mu})\;\varphi_+,\\
&&\dot\varphi_+=-\kappa_i(1-3\mu/4)\;\varphi_++\kappa_s\;\varphi_p+g_+(t),\\
&&\dot\varphi_-=-\kappa_i\mu/4\;\varphi_-+g_-(t),
\EY
where the stochastic sources are determined by the pair
correlation functions
\BY
&&\langle f_+(t)f_+(t^\prime)\rangle=2\kappa_s N\;\delta(t-t^\prime),\\
&& \langle f_-(t)f_-(t^\prime)\rangle=-2\kappa_s
N\;\delta(t-t^\prime),\\
&&\langle g_+(t)g_+(t^\prime)\rangle=-\kappa_s /(2N)\;\delta(t-t^\prime),\\
&& \langle g_-(t)g_-(t^\prime)\rangle=\kappa_s/(2
N)\;\delta(t-t^\prime).
\EY
The solution of these equations is found in the Fourier transforms
$\varepsilon_{\pm, \omega}$, $\varepsilon_{p, \omega}$,
$\varphi_{\pm, \omega}$, $\varphi_{p, \omega}$. Their variances
and correlation functions are given by:
\BY
&&(\varepsilon_{+}^2)_\omega=2N\kappa_s(\kappa_p^2+4\omega^2)/\Lambda_\varepsilon
,\L{41}\\
&&(\varepsilon_{p}^2)_\omega=2N_p(\mu_p-1)\kappa_i\kappa_s\kappa_p/\Lambda_\varepsilon
,\\
&&(\varepsilon_{p}\varepsilon_{+})_\omega=-2N\kappa_s^2\kappa_p/\Lambda_\varepsilon
,\\
&&(\varepsilon_{-}^2)_\omega=-2N\;\frac{\kappa_s}{\kappa_i^2(1-3\mu/4)^2+\omega^2
},\L{44}
\EY
and
\BY
&&(\varphi_{p}^2)_\omega=-\frac{1}{2N_p\Lambda_\varphi}(\mu_p-1)\kappa_i^2\kappa_p
,\\
&&(\varphi_{+}^2)_\omega=-\frac{1}{2N\Lambda_\varphi}\kappa_s(\kappa_p^2+4\omega^2),\\
&&(\varphi_{p}\varphi_{+})_\omega=\frac{1}{2N\Lambda_\varphi}
\kappa_i\kappa_p^2(\mu_p-1)\sqrt{1-\mu}
,\\
&&(\varphi_{-}^2)_\omega=\frac{1}{2
N}\;\frac{\kappa_s}{(\kappa_i\mu/4)^2+\omega^2 }.\L{48}
\EY
where
\BY
\Lambda_\varphi&&=\[2\omega^2-\kappa_i\kappa_p[1-3\mu/4+(\mu_p-1)\sqrt{1-\mu}\;]\]^2+\nn\\
&& +\omega^2[\kappa_p+2\kappa_i(1-3\mu/4)]^2.\\
\Lambda_\varepsilon&&=\[2\omega^2-\kappa_p\kappa_i[\mu/4+(\mu_p-1)\sqrt{1-\mu}\;]\]^2+\nn\\
&& +\omega^2(\kappa_p+\kappa_i\mu/2)^2
\EY
\newpage

\begin{figure}[h]
 \epsfxsize=70mm
\epsfbox{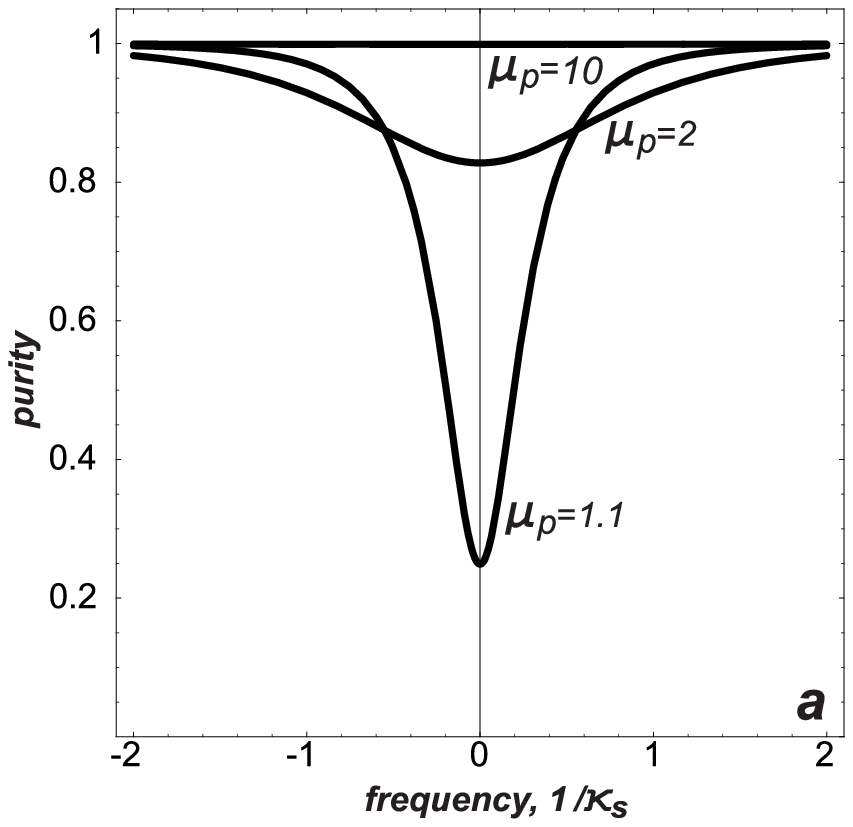}
 \epsfxsize=70mm
 \epsfbox{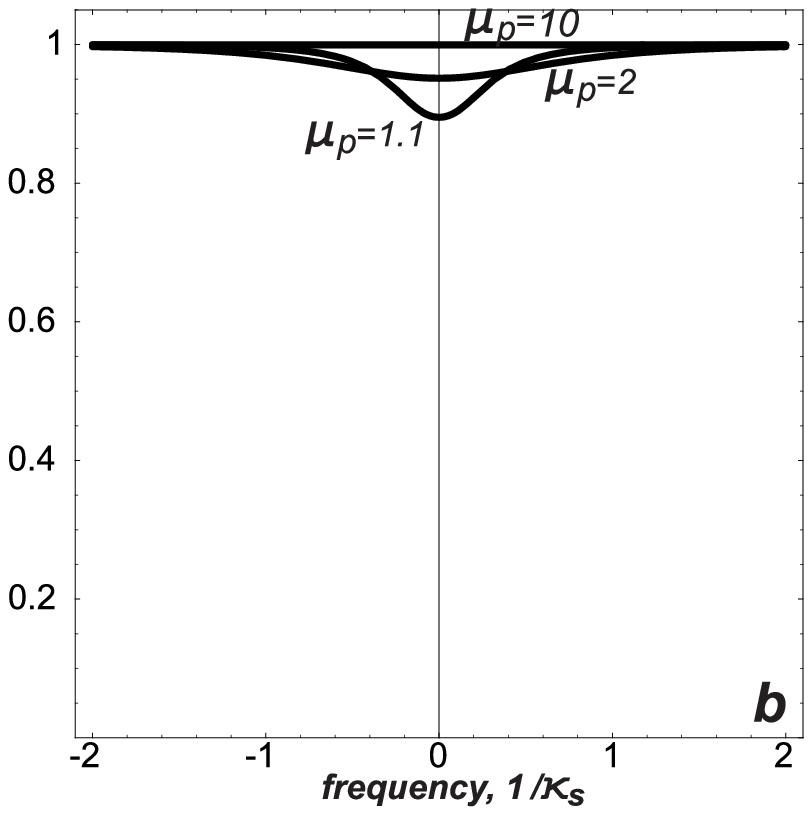}
   \caption{Purity of the output field as a function of the
     dimensionless frequency for synchronizing field parameter a)
     $\mu=0.1$; b) $\mu=0.35$ and different excesses above threshold
     for symmetry phase locking. Both positive and negative frequency
     domains are plotted, however only the positive one correspond to
     physical values}
  \label{fig:purity01}
\end{figure}

\begin{figure}[h]
 \epsfxsize=70mm
 \epsfbox{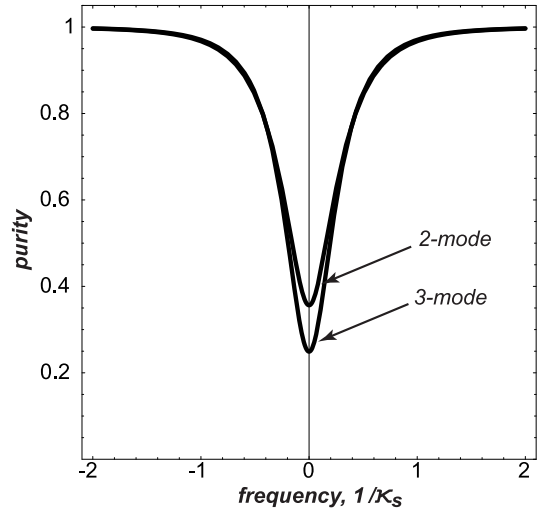}
   \caption{Partial purity of the two-mode OPO compared to the system
     purity (including the pump mode). One note that the partial
     purity is always larger the total one.}
  \label{fig:purity-partial}
\end{figure}

%
\begin{figure}[h]
\epsfxsize=70mm
 \epsfbox{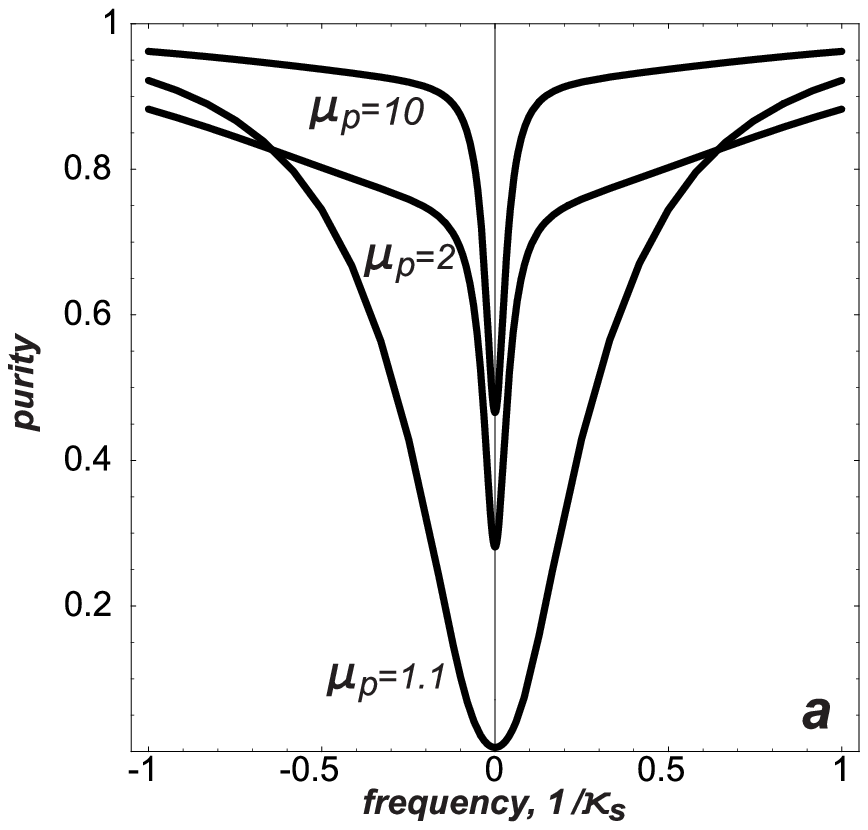}
 \epsfxsize=70mm
 \epsfbox{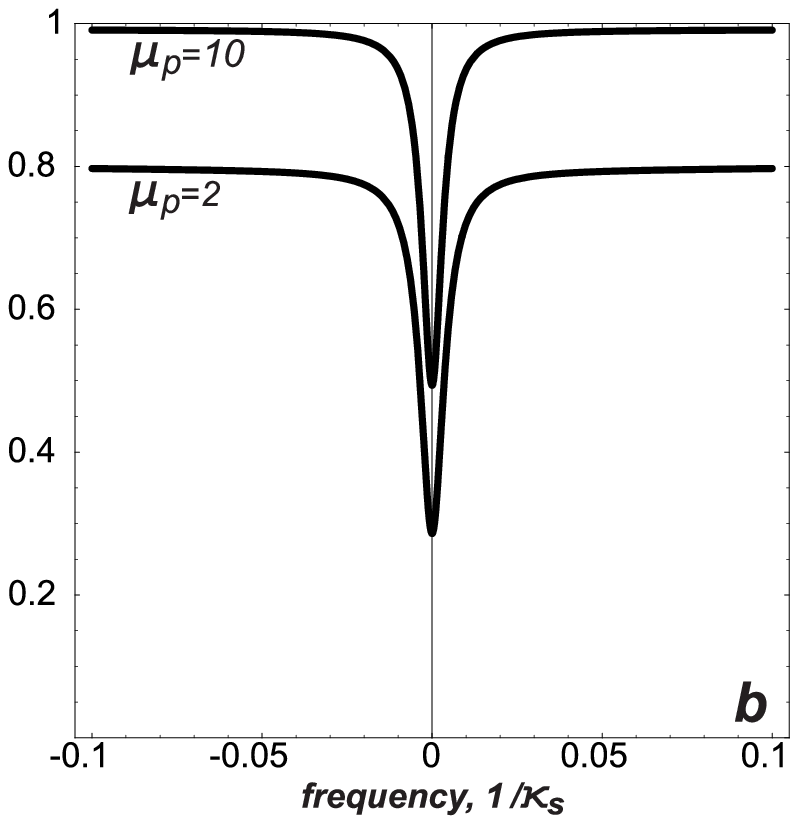}
   \caption{Purity of the output field in dependence on dimensionless frequency
  for synchronizing field parameter a) $\mu=0.1$; b) $\mu=0.01$ and different excesses above threshold
 for asymmetry phase locking}
  \label{fig:purity02}
\end{figure}
%
%
%


\begin{thebibliography}{100}

\bibitem{squeeze1} G. Milburn and D.F. Walls, Opt. Comm. {\bf 39}, 401
  (1981)
\bibitem{twin1} S. Reynaud, C. Fabre and E. Giacobino, JOSA B {\bf 4},
  1520 (1987)
\bibitem{squeeze2} L. Wu, H. Kimble, J. Hall, and H. Wu,
  Phys. Rev. Lett.  {\bf 57} 2520 (1986).
\bibitem{twin2} A. Heidmann, R. J. Horowicz, S. Reynaud, E. Giacobino,
 C. Fabre and G. Camy, Phys. Rev. Lett. {\bf 59}, 2555 (1987) 
\bibitem{squeeze3}  C Fabre, E Giacobino, A Heidmann, L Lugiato, S
  Reynaud, M Vadacchino and  Wang Kaige, Quantum Opt. {\bf 2}, 159 (1990)
\bibitem{EPR1} M.D. Reid and P.D. Drummond, Phys. Rev. Lett. {\bf 60},
  2731 (1988)
\bibitem{EPR1b} Z.Y. Ou, S.F. Pereira, H.J. Kimble and K.C. Peng,
  Phys. Rev. Lett. {\bf 68}, 3663 (1992)
\bibitem{EPR2} A. S. Villar, L. S. Cruz, K. N. Cassemiro,
  M. Martinelli, and P. Nussenzveig, Phys. Rev. Lett. {\bf 95}, 243603
  (2005)
\bibitem{pompe2} K. Kasai, G. Jiang and C. Fabre, Europhys. Lett. {\bf
  40}, 25 (1997)
\bibitem{tripartite} A. S. Villar, M. Martinelli, C. Fabre, and
  P. Nussenzveig, Phys. Rev. Lett. {\bf 97}, 140504 (2006)
\bibitem{ProgressInOptics} S. Reynaud, A. Heidmann, E. Giacobino and
  C. Fabre, Progress in Optics {\bf XXX},1 (1992) 
\bibitem{book} D.F. Walls and G.J. Milburn, {\it Quantum Optics},
  Springer Study Edition (1995)
\bibitem{Graham} R.Graham, H.Haken, {\it Z.Phys}, \textbf{210}, 276 (1968);
\textbf{210}, 319 (1968); \textbf{211}, 469 (1968).
\bibitem{conditional1} J. Laurat, T. Coudreau, N. Treps, A. Maître,
  and C. Fabre, Phys. Rev. Lett. {\bf 91}, 213601 (2003)
\bibitem{Gardiner}
K.L.McNeil, C.W.Gardiner, {\it Phys.Rev.} A, {\bf 28}(3), 1560,
(1983)
\bibitem{Sudarshan} E. Sudarshan, Phys. Rev. Lett. {\bf 10}, 277 (1963)
\bibitem{Glauber} R.J. Glauber, Phys. Rev. Lett. {\bf 10}, 84 (1963)
\bibitem{Golubev} Yu. M. Golubev and I.V. Sokolov, Sov. Phys. JETP
  {\bf 60}, 234 (1984)
\bibitem{Drummond1}
K.Dechoum, P.D.Drummond, S.Chaturvedi, M.D.Reid, "Critical
fluctuations and entanglement in the nondegenerate parametric
oscillator"
\bibitem{Gol} Yu. M. Golubev, Sov. Phys. JETP {\bf 38}, 228 (1974)
\bibitem{Armenian}
K.V.Kheruntsyan, K.G.Petrosyan, {\it Phys. Rev. A},\textbf{62},
015801 (2000);
\bibitem{Laurat2004} J. Laurat, T. Coudreau, N. Treps, A. Maître and
  C. Fabre, Phys. Rev. A {\bf 69}, 033808 (2004)
\end{thebibliography}
\end{document}